\documentclass[prc,aps,nofootinbib,showpacs,twocolumn,superscriptaddress]{revtex4}   
\usepackage{epsfig}  
\usepackage{graphicx}  
\usepackage{amssymb}
\usepackage{amsmath}
\usepackage{color}  

\usepackage{dsfont}

\renewcommand{\vec}[1]{\mbox{\boldmath $#1$}}

\begin{document}  
  
\title{Multi-dimensional fission model with a complex absorbing potential}  
  
\author{Guillaume Scamps}  
 \email{scamps@nucl.phys.tohoku.ac.jp}  
\affiliation{Department of Physics, Tohoku University, Sendai 980-8578, Japan}

 \author{Kouichi Hagino}  
 \email{hagino@nucl.phys.tohoku.ac.jp}  
 \affiliation{Department of Physics, Tohoku University, Sendai 980-8578, Japan}
\affiliation{Research Center for Electron Photon Science, Tohoku University, 1-2-1 Mikamine, Sendai 982-0826, Japan}
    
\begin{abstract}  
We study the dynamics of multi-dimensional quantum tunneling by introducing 
a complex absorbing potential to a two-dimensional model for spontaneous 
fission. We fist diagonalize the Hamiltonian with the complex potential 
to determine a resonance state 
as well as its life-time. We then solve 
the time-dependent Schr\"odinger equation with such basis 
in order to investigate the tunneling path. 
We compare this method with the semi-classical method for multi-dimensional 
tunneling with imaginary time. 
A good agreement is found both for the life-time and for the tunneling path.
\end{abstract}

 
\maketitle  

\section {Introduction}

A spontaneous fission is a typical example of multi-dimensional quantum 
tunneling, and its description has remained a challenge in nuclear theory. 
The first step 
for any calculation 
is to construct a potential energy surface in 
a multi-dimensional space. 
A standard approach for this 
is the macroscopic-microscopic method, with which the 
potential surface is constructed using the shell correction 
method of Strutinsky \cite{Brack72,Moller01,Moller09}. 
In recent years, a microscopic description based on a self-consistent 
mean-field theory has also been attempted 
\cite{Skalski08,Staszcak09,Baran11,Sadhukhan13,Mcdonnell14,Sad14,Warda12,Robledo14,Giuliani14,Goutte05,Goutte11,Afanasjev10,Lu12,Zhao15}. 
Even though a microscopic understanding of fission phenomena is important, 
there have still been many open problems to be solved. 
For instance, 
the choice of relevant degrees of freedom is still under 
discussion \cite{Sad14} and a large uncertainty may arise from a choice of 
energy functional \cite{Bur04}. 
Moreover, a difficulty in constrained mean-field 
calculations has also been pointed 
out \cite{Moller09,MS96}. 

In order to calculate the fission life-time, 
most of the calculations, both with the macroscopic and the 
microscopic approaches, rely on the semi-classical approximation. 
That is, one often searches the least action path in a multi-dimensional 
space \cite{Brack72, Sadhukhan13,Sad14} or equivalently 
solves the Newtonian equations with the 
inverted potential \cite{Sch86,KI89,I94}. 

In this paper, we investigate this problem from a different perspective. 
That is, we solve the time-dependent Schr\"odinger equation (TDSE) 
and monitor the time evolution of wave function 
in a fully quantum mechanical manner. 
This method 
provides a good intuitive description for particle decays, and has  
been applied to systems where the decay width of 
a resonance state is about the same order of magnitude 
as the resonance energy \cite{Car94,Tal99,Lac99,Mar12,OHS14}. 

In the previous applications of this method, 
because the TDSE has been solved by a numerical integration, 
the time step had to be small 
with respect to the typical time scale of the process. 
The evolution was then restricted to the small time regime.  
In this paper, we propose a new method 
introducing a complex absorbing potential in the exterior region.  
The complex absorbing potential has been 
used in time-dependent calculations \cite{Ueda03,Nak05,Rei06,Sca13,Ele14} 
in order to absorb the wave function at the boundary 
so as to avoid the reflexion that would otherwise perturb the dynamics.
For the decay problems, the absorbing potential simulates the outgoing 
wave boundary condition, which is imposed when 
one constructs a Gamow state. 
Notice that 
this method is intimately related to 
the so called complex absorbing potential method (the CAP method), which 
has been developed to compute resonances states in 
atomic physics \cite{RM93,Feu03,San01} as well as 
in nuclear physics \cite{Mas02,IY05,Ota14}, even though we do not 
take the limit of vanishing 
complex absorbing potential. 

The paper is organized as follows, 
In Sec. \ref{sec:method}, we present 
the formalism for time-dependent calculations with a complex 
absorbing potential, which 
describes quantum mechanically 
multi-dimensional 
tunneling decay problems. 
In Sec. \ref{sec:results}, we apply this method to a simple two-dimensional 
model for spontaneous fission. We compare the results with 
those in the 
semi-classical approximation both for the one- and two-dimensional 
problems. We then summarize the paper in Sec. \ref{sec:summary}. 

\section{Formalism}
\label{sec:method}

In order to investigate the multi-dimensional tunneling problem, 
we solve the TDSE in a finite box. 
To this end, 
we add an imaginary potential to the Hamiltonian $H$, that is, $H'$=$H+iW(r)$. 
The imaginary potential $iW(r)$ absorbs the outgoing flux, 
and should be applied only at the edge of the box 
in order not to perturb the physical behavior of the decay process. 
This is effectively equivalent to imposing the outgoing wave 
boundary condition for resonance states. 
The TDSE can be integrated as,
\begin{align}
	|\Psi(t) \rangle = e^{-\frac{i}{\hbar}t \hat{H}'} | \Psi_0 \rangle, 
\label{eq:propagator}
\end{align}
where $|\Psi_0 \rangle$ is the initial wave function at $t$=$0$. 
 
To compute easily the propagator in Eq. (\ref{eq:propagator}), 
we expand the wave function on 
the bi-orthogonal basis \cite{Mor52,Hussein95,HT98} 
formed by the left and right eigenfunctions of the Hamiltonian $H'$,
\begin{align}
	H' | \varphi^r_i \rangle = E_i  | \varphi^r_i \rangle \quad {\rm and} \quad   \langle \varphi^l_i | H' = E_i  \langle  \varphi^l_i | . \label{eq:eig_left_right}
\end{align}
Notice that the eigenvalues $E_i$ are complex since the Hamiltonian $H'$ 
is non-Hermitian. The bi-orthogonal basis forms the completeness 
relation as 
$\sum_i| \varphi^r_i \rangle \langle  \varphi^l_i | = \mathds{1} $, which 
leads to 
the simple evolution, 
\begin{align}
	| \Psi(t) \rangle = \sum_i e^{-\frac{i}{\hbar}t E_i}  \langle  \varphi^l_i  | \Psi_0 \rangle   
| \varphi^r_i \rangle \label{eq:TD_diag} .
\end{align}
An advantage of this method is that 
the evolution of the system can be followed for a very long time. 
This is particularly suitable for a tunneling process 
where the life-time is several orders of magnitudes longer 
than the characteristic time scale of the system.

Among the eigenstates of $H'$, 
we identify 
the state $| \varphi^r_i \rangle$ 
which has the smallest value of 
the imaginary part of eigenenergy, $E_i$=$E^r_i-i\Gamma_i/2$, 
with the physical resonance state.  
The real part of energy, $E^r_i$ corresponds to the resonance energy 
while the imaginary part $\Gamma_i$ corresponds 
to the resonance width. 
In fact, it is straightforward with the TDSE to show that this state 
has the life-time of $\tau_i$=$\hbar / \Gamma_i$. 
In this method, the details of the initial wave function 
$|\Psi_0 \rangle$ is unimportant as long as it has an appreciable overlap 
with the resonance wave function, $| \varphi^r_i \rangle$. 

\section{results}
\label{sec:results}

\subsection{Model Hamiltonian}
\label{sec:model}

We now apply the formalism presented in the previous section 
to a fission problem and compare the 
results with those in the semi-classical approximation. 
To this end, we employ a two-dimensional fission model considered 
in Refs. \cite{Bri83,Tak95}. 
This model consists of 
the elongation $R$ between the two fission fragments and an intrinsic 
degree of freedom $\xi$ coupled to the elongation. 
The Hamiltonian then reads, 
\begin{align}
H(R,\xi) &= - \frac{\hbar^2}{2M} \frac{\partial^2}{\partial R^2} + U(R) \nonumber \\
 &- \frac{\hbar^2}{2m} \frac{\partial^2}{\partial \xi^2} 
+ \frac12 m \omega^2 \xi^2 + g R \xi. 
\label{eq:H}
\end{align}
The potential $U(R)$ is chosen to be 
\begin{align}
U(R) = \frac12 M \Omega^2 R^2 \left(1-\frac{R}{R_b}\right), 
\label{eq:pot0}
\end{align}
in order to form a barrier. 
The total potential, $V(R,\xi)$=$U(R)+ \frac12 m \omega^2 \xi^2 + g R \xi$, 
has a saddle at 
\begin{eqnarray}
R_s&=&\frac{2R_b}{3M\Omega^2}\left(M\Omega^2-\frac{g^2}{m\omega^2}\right), \\
\xi_s&=&-\frac{2gR_b}{3Mm\Omega^2\omega^2}
\left(M\Omega^2-\frac{g^2}{m\omega^2}\right), 
\end{eqnarray}
with the barrier height of 
\begin{equation}
V_b=\frac16\left(\frac{2R_b}{3M\Omega^2}\right)^2\left(M\Omega^2-\frac{g^2}{m\omega^2}\right)^3.
\end{equation}
When the intrinsic degree of freedom $\xi$ is neglected, the saddle is
at $R_s$=$2R_b/3$ with the height of $V_b$=$2M\Omega^2R_b^2/27$ \cite{Bri83}.

In the calculations presented below, we use the same parameters as those 
in Refs. \cite{Bri83,Tak95} except for $R_b$, which we vary to study 
the tunneling in the potential with different barrier heights. 
Those parameters were determined in order to mimic the 
symmetric fission of $^{234}$U with a coupling to the beta vibration. 
The parameters are then taken to be 
$\hbar \Omega$=$\hbar \omega $=$ 0.97$~MeV, $g^2$=$Mm \Omega^2/(16\hbar^2)$, 
$M$=$234M_N/4$   and $ m$=$3A M_N R_0^2/(8\pi)$, where $M_N$ is the nucleon 
mass, $A$=$234$ is the atomic number of the nucleus, 
and $R_0$=$1.2 A^{1/3}$ fm is the equivalent sharp radius (notice that 
the vertical axis in Fig. 4 in Ref. \cite{Tak95} is actually $R_0\xi$, rather 
than $\xi$ itself). 
In the actual calculations, we modify the total potential 
to a constant value in the outer region in order to avoid the divergence 
of the potential, that is, $V(R,\xi)\to \max(V(R,\xi),-3)$.
%

\subsection{1-Dimensional Problem}
\label{sec:1D}

Before we discuss the tunneling dynamics in the two-dimensional space,
let us first solve the problem in one dimension neglecting the $\xi$
degree of freedom. In the one-dimensional problem,
the tunneling path is trivial,
and the decay life-time is obtained in the semi-classical approximation as, 
\begin{align}
  \tau = \frac{2 \pi}{\Omega}\,e^{2 S/\hbar},
\label{eq:WKB_1D}
\end{align}
with
\begin{align}
  S=\int^{R_1}_{R_0}\sqrt{2 M (U(R)-E_0)}\,dR,
\end{align}
where $E_0$ is the energy of the decay state, and $R_0$ and $R_1$ are
the inner and the outer turning points, respectively, satisfying
$U(R_0)$=$U(R_1)$=$E_0$.
For a cubic potential given by Eq. (\ref{eq:pot0}), the semi-classical
formula can also be transformed to \cite{Sch86,Bri83},
\begin{align}
  \tau = \frac{1}{\Omega}\,\sqrt{\frac{\pi \hbar}{60 S'}}\,e^{2 S'/\hbar},
\label{eq:WKB}
\end{align}
with
\begin{align}
  S'=\int^{R_1'}_0\sqrt{2 M U(R)}\,dR,
\end{align}
where $U(R_1')=U(0)=0$.

\begin{figure}[!ht]   
	\centering\includegraphics[width=\linewidth]{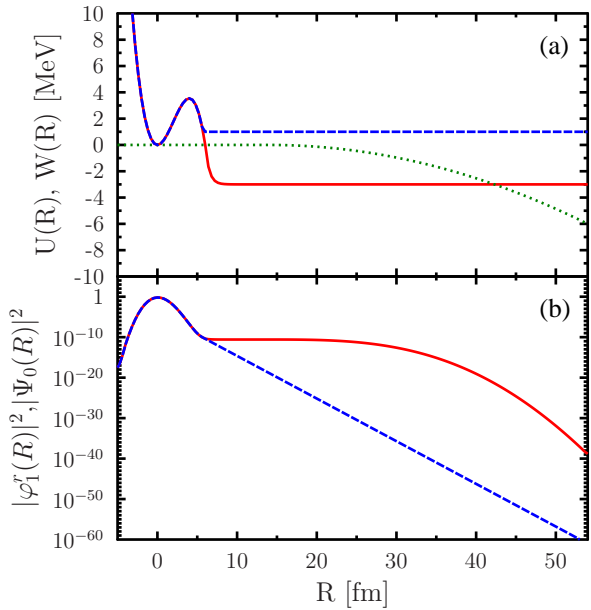}  
	\caption{(Color online)  (a) The one dimensional potential $U(R)$
          with $R_b$ = 6 fm (the red solid line). The figure also shows the 
          modified potential for the initial wave function (the blue
          dashed line) and the absorbing potential $W(R)$ (the green
          dotted line).
          (b) The resultant initial wave function (the blue dashed line) and the
          resonance wave function (the red solid line).} 
	\label{fig:res_pot_wave_2} 
\end{figure}

In order to solve the same problem quantum mechanically,
we first determine the initial wave function $|\Psi_0 \rangle $ by
modifying the potential $U(R)$ 
so that the modified potential has a bound state.
To this end, we replace the potential outside the barrier
with a constant value \cite{OHS14,GSNV04}, which we take 1 MeV 
as shown by the dashed line 
in Fig. \ref{fig:res_pot_wave_2} (a). 
The corresponding initial wave function is plotted by the dashed line
in the lower panel of Fig. \ref{fig:res_pot_wave_2}. 

For the complex absorbing potential, 
we employ 
the shifted polynomial function \cite{IY05,Ota14},
\begin{align}
  W(R) = W_0  (R-R_a)^2\theta(R-R_{a}), \label{eq:potim}
\end{align}
with $W_0$=$-0.387$~MeV and $R_a$=$14$~fm.
We have confirmed that the results do not significantly change even
if we vary the value of $W_0$ and $R_a$ as well as the size of the box.

With this complex potential, 
we then
determine the ensemble of eigenstates and eigenvalues
of the Hamiltonian $H'$ 
defined by Eq. \eqref{eq:eig_left_right}.
It should be mentioned here that due to the large difference in the 
order of magnitude between the real and imaginary parts of the resonance
energy, it is numerical necessary to use the 
quadrupole precision in the program. 
With this prescription
physical quantities can be calculated up to about 34 decimal digits.  
This allows us to calculate a life-time of the order 
of billion of years when the characteristic time of the system
is of the order of $10^{-22}$~s.
Another important parameter in the calculations
is the lattice mesh size $\Delta R$, that has to be small enough
in order to describe correctly the tunneling wave function.
In our calculations, we take 
$\Delta R$=$0.25$~fm 
with the
finite difference formula with 9 points for
the second derivative in the kinetic energy operator.

To select the physical resonance wave function
among the eigenstates of $H'$, 
we take the lowest energy
state $|\varphi_1^r\rangle$
which has the maximum overlap with the initial
state. 
The resulting wave function is shown in Fig. \ref{fig:res_pot_wave_2} (b) by
the solid line.
We see that the wave function is smoothly damped up to a
factor of $10^{30}$ before reaching the reflecting edge
of the lattice at $R$=54~fm.
The decay width can then be read off 
from the imaginary part of the eigenenergy. 

\begin{figure}[!ht]   
	\centering\includegraphics[width=\linewidth]{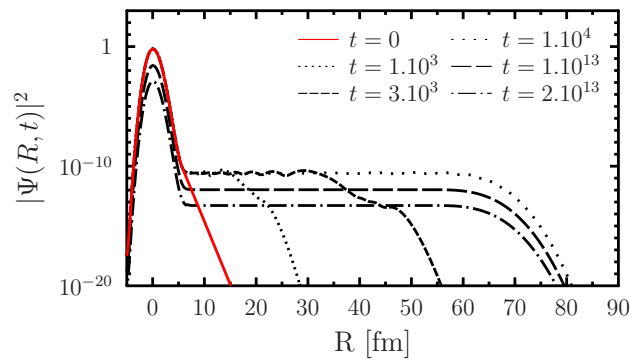}  
	\caption{(Color online)  The time-evolution of the
          square of the wave function during the decay process. The
          time is indicated in unit of fm/$c$.} 
	\label{fig:film_1D} 
\end{figure} 

In order to have a more intuitive picture on the decay process,
we compute the time-evolution of the wave function according to
Eq. \eqref{eq:TD_diag}.
The probability density at different times
obtained with a 
larger value of $R_a$ ($R_a$=54 fm) 
is shown in Fig. \ref{fig:film_1D}.
At $t$=$0$, the confining in the modified potential is suddenly
removed,
and the initial wave function is coupled to the continuum. 
In the first instant of the dynamics,
one can see the emitted wave going outward the potential barrier. 
After about $t$=$10^4$ fm/$c$,
the emission becomes stationary,
and the wave packet has the same shape as the metastable
wave function constructed by diagonalizing the Hamiltonian (see the solid line in
Fig. \ref{fig:res_pot_wave_2} (b)).
At later time, the wave packet is absorbed exponentially
keeping the same spatial shape.

\begin{figure}[!ht]   
	\centering\includegraphics[width=\linewidth]{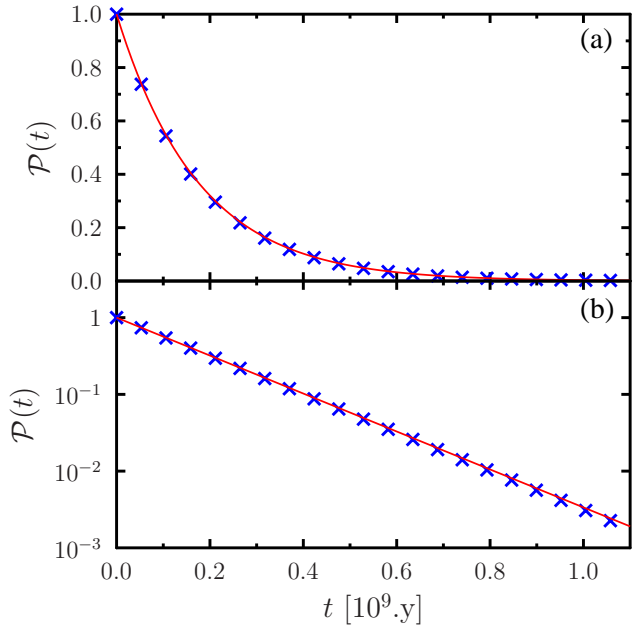}  
	\caption{(Color online)
          The survival probability as a function of time computed dynamically
          using the formula \eqref{eq:surv} with
          $R_{\rm lim}$=20~fm (the blue crosses). 
The $R_b$ parameter in the potential is taken to be 
          $R_b$=11~fm. The survival probability  
          is plotted both in the linear scale (the upper panel) and in the
          logarithmic scale (the lower panel).
          For a comparison, the figure also shows the survival probabilities 
          for an exponential decay 
          with the decay width obtained from the semi-classical
          approximation given by Eq. \eqref{eq:WKB_1D}
          (the red solid line). } 
	\label{fig:decr} 
\end{figure}

The decay width can be calculated with the TDSE
by computing the survival probability defined as 
\begin{align}
{\cal P}(t) = \int_{- \infty}^{R_{\rm lim}} | \Psi(R,t)|^2 dR, \label{eq:surv}
\end{align}
as a function of time, where
$R_{\rm lim}$ is taken outside the barrier. 
The survival probability
obtained with $R_{\rm lim}$=20~fm 
is plotted in Fig. \ref{fig:decr}
for the potential with $R_b$=11~fm.
We find that both the TDSE method and the imaginary part of the eigenenergy
of $H'$ yield $1/\tau$=$5.76\times 10^{-9}$ (1/year) while the
semi-classical approximation yields
$1/\tau $=$ 5.69\times 10^{-9}$ (1/year). 
The agreement between the quantal and the semi-classical
calculations is rather good for this parameter set,
partly because the barrier is high and the multiple reflections under
the barrier can be neglected.

\subsection{2-Dimensional Problem}
\label{sec:2D}

\begin{figure}[!ht]   
	\centering\includegraphics[width=\linewidth]{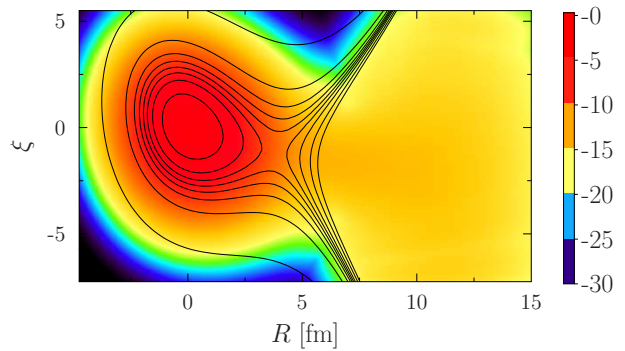}  
	\caption{(Color online) The wave function for the meta-stable
          state plotted in the logarithmic scale. 
          This state corresponds to an eigenstate of the Hamiltonian $H'$
          for the potential with $R_b$=6 fm and with the
          imaginary potential given by Eq. \eqref{eq:potim} with
          $W_0$=$-0.1$~MeV and $R_a$=$11$~fm. The total potential $V(R,\xi)$
          is also plotted by the contour lines.}
	\label{fig:wave_H} 
\end{figure}

We now discuss the tunneling dynamics in the two-dimensional
surface.
To this end,
we diagonalize the Hamiltonian $H'$ 
in 2 dimensions in a lattice of dimension
$R\in[-4.75$ fm $ : 16$ fm$]$ and $\xi \in [-7.25: 5.5 ]$
with a mesh size of $\Delta R$=0.25~fm and $\Delta \xi$=0.25.
The imaginary potential is implemented with the same expression
as Eq. \eqref{eq:potim} with $W_0$=$-0.1$~MeV and $R_a$=$11$~fm. 
The reference wave function to be used to select
the physical resonance state 
is obtained in a similar manner as in the
previous subsection. 
The wave function for the metastable state is 
plotted in Fig.~\ref{fig:wave_H} for the choice of
$R_b$=$6$ fm, together with the total potential
$V(R,\xi)$ in the contour lines.
One can see that this wave function is well confined inside the
barrier with a small component outside due to the quantum tunneling,
as in the one-dimensional case shown in Fig. 
\ref{fig:res_pot_wave_2}. 

\begin{figure}[!ht]   
	\centering\includegraphics[width=\linewidth]{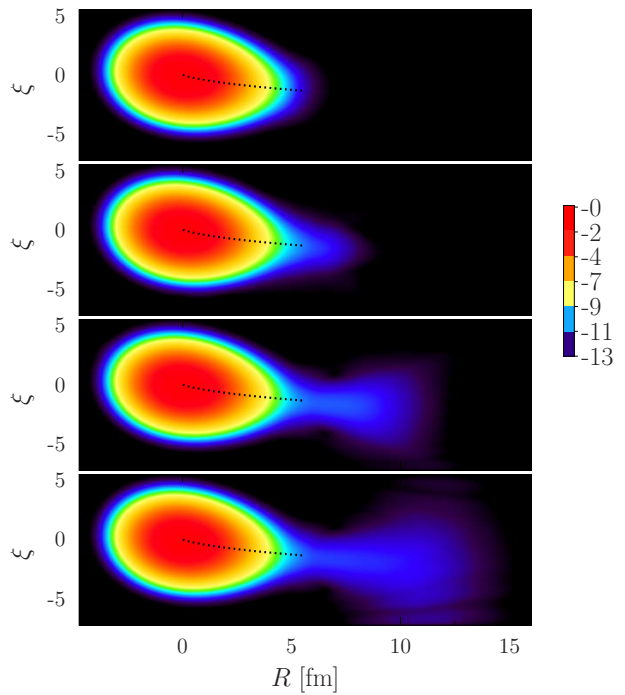}  
	\caption{(Color online) The time evolution of the two-dimensional
          wave function at time $t$=0, $t$=250 fm/$c$, $t$=500 fm/$c$
          and $t$=1250 fm/$c$. For a comparison, the semi-classical tunneling
          path is also shown with the dashed line.}
	\label{fig:film} 
\end{figure}

Notice that 
this wave function corresponds to the wave function after
the decay becomes stationary. 
The pre-stationary decay at the first instant of the decay process 
is shown in Fig. \ref{fig:film},
following the time evolution from the reference wave function as
the initial state. 
One can see that the flow occurs in a small region in the potential energy
surface. 

We compare the quantum mechanical flow with the semi-classical tunneling
path. 
To this end, we follow the method described in Refs. \cite{Sch86,KI89,I94}.
That is, the semi-classical tunneling path is determined by
solving the classical equations of motion
in an inverted potential, $ V(R,\xi) \rightarrow - V(R,\xi)$.
For the initial condition, 
one can use $R(0)$=$\epsilon\sqrt{\frac{\hbar}{2M\Omega}}\,
\cos\theta$, $\xi(0)$=$
\epsilon\sqrt{\frac{\hbar}{2m\omega}}\,
\sin\theta$, and $\dot{R}(0)=\dot{\xi(0)}$=$0$, where $\epsilon$ is
a small number. By searching with $\theta$ from 0 to 2$\pi$, one finds
a special angle $\theta$ for which the classical path reaches
the equi-potential surface (and bounces back to the origin if one continues
to follow the time evolution of the path), while for other
values of $\theta$ the classical path is reflected before it reaches the
equi-potential surface \cite{Sch86,KI89,I94,Tak95,Brink88}.
  This special path is referred to as the
escape path, and plays an important role in the semi-classical theory
of multi-dimensional quantum tunneling.
The escape path so obtained is denoted by the dashed line in
Fig. \ref{fig:film} (there is only one escape path for the potential
considered in this paper).
It is remarkable that the quantum
mechanical time evolution almost follows the semi-classical path.
We have confirmed that the tunneling path obtained by minimizing
the classical action with the algorithm in Ref. \cite{Bar81}
provides the same result.

\begin{figure}[!ht]   
	\centering\includegraphics[width=\linewidth]{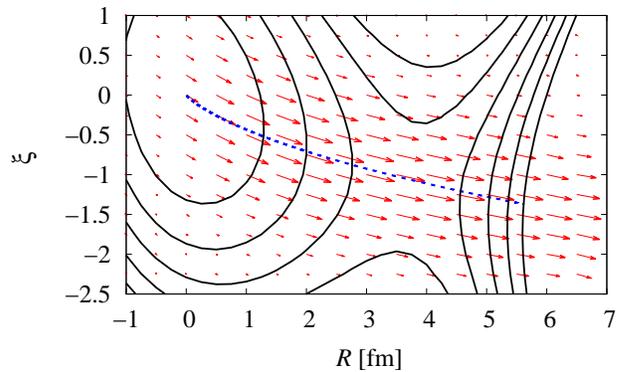}  
	\caption{(Color online) 
	  A comparison of the quantum mechanical flux computed with
          Eqs. \eqref{eq:flow-R} and \eqref{eq:flow-xi} (the red arrows)
          with the semi-classical escape path (the blue dashed line).
          The two-dimensional potential energy surface is also shown by the
          black contour lines.
	} 
	\label{fig:vector} 
\end{figure}

To make a further comparison of the tunneling path,
we compute the flux, $\vec{j}$=$(j_R,j_\xi)$,  from the resonance wave 
function as, 
    \begin{align}
      j_R &= \frac{\hbar}{2iM} \left( \varphi^{r*}_1 \,\frac{\partial\varphi^{r}_1}
      {\partial R}  -  \varphi^{r}_1 \,\frac{\partial\varphi^{r*}_1}{\partial R}
            \right),        \label{eq:flow-R}
   \end{align}
and 
    \begin{align}
      j_\xi &= \frac{\hbar}{2im} \left( \varphi^{r*}_1 \,\frac{\partial\varphi^{r}_1}
      {\partial \xi}  -  \varphi^{r}_1 \,\frac{\partial\varphi^{r*}_1}{\partial \xi}
            \right).        \label{eq:flow-xi}
   \end{align}
The flux is shown in Fig.~\ref{fig:vector} and is
compared to the semi-classical escape path.
The quantum flow is a collection of
all the trajectories connecting the
region around the origin with the continuum region.
We see that the main quantum mechanical flow 
is systematically
parallel to 
the semi-classic trajectory.  
The semi-classical path can thus be regarded as the mean trajectory
of the quantum flow, where as the quantum flow takes into account
the quantum mechanical fluctuation around the classical trajectory.

\begin{figure}[!ht]   
	\centering\includegraphics[width=\linewidth]{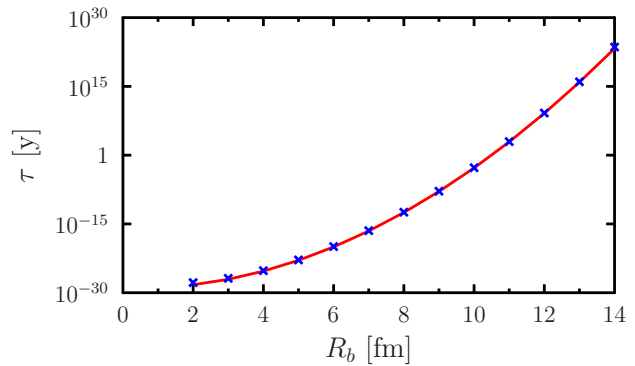}  
	\caption{(Color online) The fission life-time
          as a function of the $R_b$ parameter in the potential $U(R)$.
          The life-time computed quantum mechanically is denoted by the
          blue crosses, while that evaluated with the semi-classical method
          using Eq. \eqref{eq:WKB_2D} is denoted by the red line. } 
	\label{fig:res_2D} 
\end{figure}

In addition to the agreement for the tunneling path,
we also compare in Fig.~\ref{fig:res_2D}
the resulting fission life-time.
Varying the $R_b$ parameter in the potential $U(R)$,
we obtain a range of fission life-time from 
the characteristic time of the system to a life-time longer 
than the actual life-time of the $^{238}$U nucleus.
The quantum mechanical life-time is computed from the imaginary
part of the resonance energy, while the life-time in
the semi-classical approximation is evaluated using the formula
Eq. (\ref{eq:WKB}) with the action integral evaluated along the
escape path $\cal P$ in the imaginary time,
\begin{align}
  S'&=\int_{\cal P}\left(M \dot R^2 + m \dot \xi^2\right)\,d\tau  \label{eq:WKB_2D}.
\end{align}
The latter is equivalent to approximating the pre-exponential factor
by that for a one-dimensional problem with a cubic
potential along the escape path \cite{Tak95}.
In the range shown in the figure,
a good agreement is
found between the two methods, the maximum deviation being
up to about 20\%. 
Evidently, the semi-classical method 
provides a good approximation to the multi-dimensional 
tunneling problem for this
Hamiltonian.

\section{Summary}
\label{sec:summary}

We have presented a full-quantum method to study a decay life-time
as well as the tunneling dynamics. 
To this end, we have introduced a complex absorbing potential, 
which is effectively equivalent to the outgoing boundary condition, 
and constructed the bi-orthogonal basis. 
We have shown that this method provides a good tool to follow the time
evolution of a system over a very long time.
This enables one to compute the decay life-time from a few fm/$c$ to
the order of billion of years.
A comparison with the semi-classical approximation
has shown a good agreement between the two methods.
It has been shown that the average of 
fissioning flux in a multi-dimensional plane
corresponds to the semi-classical tunneling path. 
For the decay life-time, the two methods yield
similar values to one another, where the maximum difference is
only about by 20\%.

The quantum mechanical method discussed in this paper
provides a good alternative to the semi-classical method. It provides an
intuitive picture of multi-dimensional quantum tunneling, including the
quantum mechanical fluctuation of the classical path. 
It will also provide a convenient method when a bifurcation of
tunneling path is important, {\it e.g.,} in the presence of a competition
of several fission modes.
We plan to apply this method, both for dynamics and a determination
of resonance state,
to more realistic systems 
using the constrained mean field theory for
the potential energy surface and/or the
time-dependent generator coordinate method \cite{Goutte05,Goutte11,PG83}.
We will report it in a separate publication. 

%
 
 \section*{Acknowledgement}
 
 G.S. acknowledges the Japan Society for the Promotion of Science
 for the JSPS postdoctoral fellowship for foreign researchers.
 This work was supported by Grant-in-Aid for JSPS Fellows No. 14F04769.

\end{document}